\documentclass{emulateapj}
\usepackage{times}
\usepackage{graphicx}

\usepackage[caption=false]{subfig}

\slugcomment{Submitted to ApJ}

\shorttitle{1FGL J1018.6$-$5856}
\shortauthors{Waisberg et al.}

\begin{document}

\title{Echelle Spectroscopy of $\gamma$-ray binary 1FGL J1018.6$-$5856}

\author{Idel R. Waisberg\altaffilmark{1} and Roger W. Romani\altaffilmark{1}
}
\altaffiltext{1}{Department of Physics, Stanford University, Stanford, CA 94305-4060,
 USA; rwr@astro.stanford.edu, idelw@stanford.edu}


\begin{abstract}

	We observed the {\it Fermi}-discovered $\gamma$-ray binary 1FGL J1018.6$-$5856 at 20 epochs over 50 days using 
the CHIRON spectrograph, obtaining spectra at $R\sim$25,000 covering $\lambda\lambda 4090-8908$\AA.
The average spectrum confirms an O6 V((f)) spectral type and extinction $E(B-V) = 1.35\pm 0.04$. 
Variable absorption line equivalent widths suggest substantial contamination by wind line features.
The limited S/N ratio hindered accurate continuum definition and prevented measurement of
a high quality radial velocity curve. Nevertheless, the best data indicate
a radial velocity amplitude $\le 40 {\rm km~s^{-1}}$ for the \ion{He}{2} lines and substantially lower
for H~I. We argue that this indicates a most likely compact object mass  $<2.2M_\odot$. While
black hole solutions are not excluded, a neutron star source of the $\gamma$-ray emission
seems preferred.
\end{abstract}

\keywords{gamma rays: stars --- binaries: spectroscopic}

\section{Introduction}

	To date only five binaries with high mass secondaries have been detected in the GeV 
band; 1FGL 1018.6$-$5856 is the only one whose binary period was discovered 
at these energies \citep{cet11,acket12}. The nature of the $\gamma$-producing compact object
in these binaries is unclear: it might be a jet-producing black hole, in analogy to $\gamma$-loud blazars
(the `$\mu$QSO model'), or the $\gamma$-rays may be generated by the magnetosphere
or wind termination shock of an energetic pulsar (the `PSR/PWN model'). This latter model certainly
applies to the PSR B1259$-$63 system, whose GeV emission is produced via Inverse Compton
scattering off wind shock particles when the PWN is compressed near periastron. For 
1FGL 1018.6$-$5856 and the other three systems (LS I+$61^o$303, LS 5039 and  Cygnus X-3),
the identity of the compact object is presently unknown. Discovery of $\gamma$-ray pulsations
would certainly indicate a neutron star primary. However, the poorly known acceleration in the binaries
makes $\gamma$-ray pulse searches prohibitively expensive and the dense plasma produced by the
massive secondary winds frustrates radio pulse searches. An alternative approach is to
constrain the compact object mass via measurement of the secondary's radial velocity. Although
the high mass and strong winds of the secondary also make this very challenging, even crude 
constraints can be useful; for example a minimum companion mass $> 3M_\odot$ would certainly
indicate a black hole. 

	After discovery of a 16.5\,d $\gamma$-ray modulation \citep{cet11}, the associated $Swift$ X-ray 
source was identified with a bright optical counterpart. A low resolution spectrum with the SAAO 1.9\,m
shows an O6V((f)) companion with an estimated mass $\approx 20-30M_\odot$ and distance $\approx 5$\,kpc
\citep{acket12}.
This system is also a radio source and  $Swift$ XRT (An et al 2013) and ATCA monitoring detect the orbital
modulation in the X-ray and radio band, respectively. Optical photometric monitoring
detects no variation, although present constraints are weak.
These properties are similar to those of the more compact ($P_b$=3.9\,d) $\gamma$-ray binary
LS 5039.  There spectroscopic and photometric studies \citep[e.g.][]{set11} 
indicate an O6.5V((f)) secondary, a substantial eccentricity $e=0.24\pm0.08$, and a modest mass function
$f(m) = 0.0049\pm 0.0006 M_\odot$. The limit on LS 5039's photometric variability, 2mmag, implies 
a high system inclination and/or a small compact object mass. Since the inclination is poorly
determined, they can only conclude that $M_1>1.8M_\odot$, allowing both black hole and neutron star
solutions.  

	In this paper we describe a project to study the radial velocity of 1FGL 1018.6$-$5856 using 
echelle spectroscopy. We adopt the orbital period estimate of \citet{colet14} $P_b= 16.549\pm0.007$\,d
with epoch MJD $55403.3 \pm 0.4$ corresponding to the $\gamma$-ray peak described in \citet{acket12}.
An independent {\it Swift/Fermi} analysis (An, priv. comm.) confirms this ephemeris. While our
results appear to show systematic variation with orbital phase, it is unclear if we have isolated
the true radial velocity curve. Nevertheless, our measurements improve the characterization of the
companion star, limit the amplitude of its motion and point to the sort of observations needed
for a definitive study.

\section{Observations and Calibrations}

	The 16.5\,d binary period is particularly inconvenient, being too long to cover in
a dedicated run, but too short for effective coverage with ad hoc $\sim$ monthly observations.
The best solution was queue observing and we were awarded 25h of queue time
(NOAO-11B-0092) on the CTIO 1.5\,m using the CHIRON fiber-coupled echelle spectrograph \citep{sch10}. 
1FGL 1018.6$-$5856 
is relatively faint for CHIRON at $V = 12.5$, so CHIRON was run in the slitless 
(2.5$^{\prime\prime}$ fiber aperture) mode for an effective resolution $R\sim 25,000$. The CCDs were 
binned $4\times 4$\, to maximize S/N and minimize readout time.
The spectra covered 4090$-$8908\AA , with substantial order overlap except in the reddest orders.

Observations were attempted roughly every other night from March 02, 2012 (JD 2455989) to
April 20, 2012 (JD 2456022). Four nights were lost to system failures and weather; spectra were obtained
on 20 nights (Table 1). We requested 3$\times$1000\,s exposure on 1FGL 1018.6$-$5856, one night had only
a single exposure.  For each epoch we obtained ThAr arc exposures immediately before the target observations
for wavelength calibration. For the first epoch only we also observed the flux standard
HD60753. CHIRON was undergoing final commissioning during these observations, and the obtained
S/N was substantially lower than predicted by the design plots (predicted peak S/N/extracted pixel $\sim$ 40 in 
each 1000s exposure). In practice, we obtained S/N/extracted pixel $\sim$ 25 at the order peaks
in our best exposures (in line with recent revised 6.5\% instrument efficiency).  More typical
S/N/extracted pixel was $\approx 17$, with a few epochs much worse.  We were not able to ascertain from 
the night logs why the S/N on source varied so dramatically, although from the low apparent
flux at some epochs, we suspect that poor transparency and imperfect fiber acquisition 
may play a role. In the end we analyzed the 17 epochs 
with S/N$>15$ in the summed, combined spectra. These spectra sample four binary orbits.

\begin{deluxetable}{lrrl}[h!!]
\tablecaption{\label{Journal} CHIRON Observations}
\tablehead{
\colhead{Date} & \colhead{Nexp} & \colhead{S/N$_{6000{\rm \AA}}$} &\colhead{$\phi_B$}
}
\startdata
%
03/02 & 3 & 36 & 0.41 \cr
03/06 & 1 & 24 & 0.65 \cr
03/10 & 3 & 12 & 0.89 \cr
03/14 & 3 & 33 & 0.13 \cr
03/16 & 3 & 32 & 0.25 \cr
03/18 & 3 & 44 & 0.37 \cr
03/20 & 3 & 37 & 0.49 \cr
03/22 & 3 & 33 & 0.61 \cr
03/24 & 3 &  3 & 0.73 \cr
03/26 & 3 & 31 & 0.85 \cr
03/28 & 3 & 18 & 0.98 \cr
03/30 & 3 & 18 & 0.09 \cr
04/02 & 3 & 28 & 0.28 \cr
04/04 & 3 & 31 & 0.40 \cr
04/06 & 3 & 21 & 0.51 \cr
04/08 & 3 &  8 & 0.63 \cr
04/10 & 3 & 22 & 0.76 \cr
04/12 & 3 & 20 & 0.87 \cr
04/18 & 3 & 30 & 0.24 \cr
04/20 & 3 & 29 & 0.38 \cr
\enddata
\end{deluxetable}

	Although the CHIRON data distribution included a standard pipeline extraction of the echelle
spectra, this was optimized for the relatively red target stars of the instrument's principal
(radial velocity planet search) program. This extraction covered $\lambda\lambda 4504-8908$\AA\
(62 orders); for our blue 06V target 14 additional orders were measurable, extending the spectral
coverage to 4090\AA.

	Standard CCD processing was performed with the IRAF package. A master bias was assembled 
and subtracted from the target, flat and lamp images. The spectra for each epoch cover 3200s.
As this is only 0.2\% of the orbital period, intra-epoch velocity smearing was negligible and the
three target exposures for each night were median combined with average $\sigma$-clipping to increase 
S/N and reject cosmic ray events. Combined flat field images were also prepared for each night.
These were used to define the apertures for each order. Sufficient inter-order pixels were available
to define flanking `background' windows for each order. A linear fit to these backgrounds
was used as a local scattered light subtraction. All extractions were performed using variance weighting, 
modeled by the CCD read noise ($4.5e^-$) and gain ($1.3 e^-$/ADU).

\begin{figure}[h!!]
\vskip 8.9truecm
\includegraphics{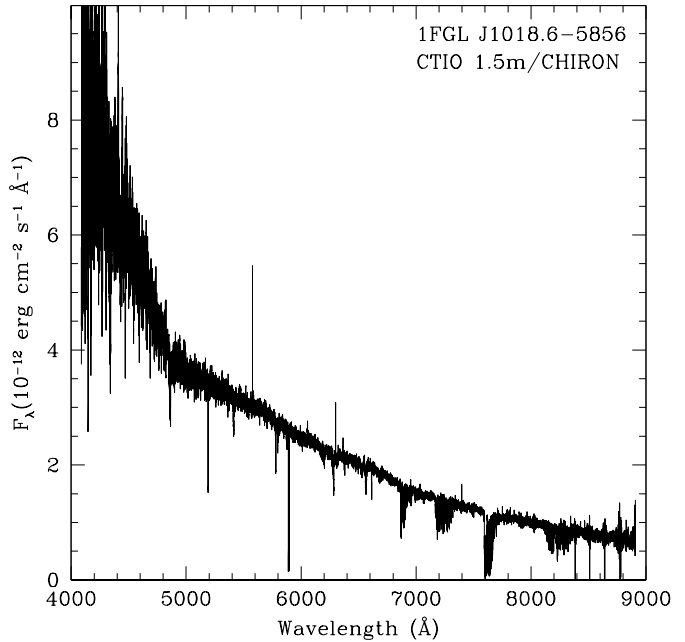}
\begin{center}
\caption{\label{fluxed} 
Fluxed CHIRON spectrum from March 02, 2012, corrected for atmospheric and interstellar
($E_{\rm B-V}=1.35$) extinction. Telluric features have not been removed.
The strongest night sky and absorption lines
are visible, along with inter-order gaps for the reddest five orders.
}
\end{center}
\vskip -0.4truecm
\end{figure}

	The extracted science spectra for each order were divided by the extracted flat spectrum.
This removed the blaze function and summed pixel-to-pixel variations, while maintaining
a relatively stable response from the smooth lamp spectrum. In principle, the true response could
be determined from the standard star spectrum. In practice, for the queue-selected standard the
tabulated fluxes were derived from a spectral model and were too sparse (10\AA\ grid) to fit an accurate
sensitivity function for each order. Accordingly we used the lamp-flattened spectra directly.
The wavelength calibration was determined from the nightly ThAr spectra. Non-linear solutions
were fit, with a typical rms residual 0.01\AA.  
Note that although CHIRON is designed to be quite stable, night-to-night shifts in the spectrum 
position on the chip were significant. Aperture migrations (1-2 pixels) were treated by
re-centering orders on the flat before extraction (and importing these traces to the 
arc and science spectra). Shifts along the dispersion direction ($\sim 0.1$\AA$\sim 5-10 {\rm km\, s^{-1}}$)
were removed by running IRAF task $\textit{re-identify}$ on each night's ThAr arcs to derive
the wavelength calibration.

	After trimming very low S/N data at the edges of each echelle order, we merged
all orders to a single spectrum, average-combining the overlap regions (figure \ref{fluxed}). Given our relatively
modest S/N for most of the spectra, this procedure generally produced an adequate match between orders.
For some epochs, small flux gradients across orders and poor inter-order match were however present. 
Such imperfections can be attributed to differences in illumination paths of and/or light path shifts
between flat and science images.  These may be removed by computation of a smoothly varying 
residual blaze function \citep{skodaet08}.
Although we attempted such corrections, the S/N of our targets and standard exposures did not allow
significant improvement. Finally, we divided the merged spectra by a low order function to
remove the global lamp response and create normalized spectra appropriate to our kinematic/line study.
The S/N weighted combined normalized spectrum from all epochs is shown in figure \ref{normspec}.
An absolute fluxed spectrum was computed from the first night's data where the adjacent
standard was used to normalize the fluxes. 
For each epoch we corrected the normalized spectrum to the heliocenter using the IRAF
routines $\textit{rvcorrect}$ and $\textit{dopcor}$.

\section{Spectrum Analysis}

\begin{figure}[t!!]
\vskip 8.9truecm
\includegraphics{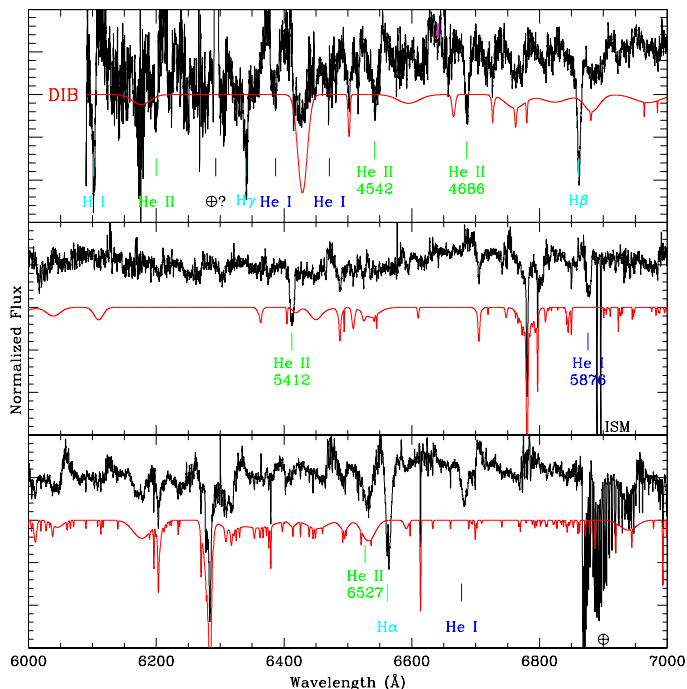}
\begin{center}
\caption{\label{normspec} 
Normalized average spectrum. Each sub-panel covers 1000\AA . The strongest lines used for the
radial velocity analysis are labeled, additional lines from these species are marked.
The synthesized DIB transmission spectrum is generated from tables in \citet{jenet96} 
(normalized to $E_{B-V}=1.35$) and is plotted offset by -0.1.
}
\end{center}
\vskip -0.5truecm
\end{figure}

	The overall spectrum (figure \ref{fluxed}) is quite consistent with the SAAO spectrum of 
\citet{acket12}, with \ion{H}{1}, \ion{He}{1} and \ion{He}{2} lines competing with strong interstellar absorption. The
\ion{He}{2}$\lambda$4541/\ion{He}{1}$\lambda$4471 ratio is a good diagnostic of sub-class for O-type stars
\citep{gray09}. From our combined spectrum we measure $2.3\pm 0.2$ for this ratio, indicating 
O6 \citep{jj95}. The \ion{N}{3} emission line triplet 4634/4640/4642 is not present, or even weakly
in absorption. Even higher S/N is needed to measure the state of this line. Its weakness
argues for a dwarf classification; overall 
the designation O6 V((f)) is supported by this spectrum. The photospheric lines are broad,
reaching FWHM 350 ${\rm km\,s^{-1}}$. There were no sky fibers for sky subtraction and so a
few of the brightest night sky lines (e.g. 5577\AA\ ) are visible. A broad emission feature
at 4293\AA\ appears in most of the spectra, with highly variable flux and high equivalent
width even when the target was weak. As this corresponds to no strong astronomical line, we
suspect a spectrograph artifact from local light. In figure \ref{normspec} we label the strongest
photospheric features.

\subsection{Interstellar Absorption}

	Our measurement of the photospheric features is complicated by particularly
strong interstellar absorption. Atomic features (e.g. Mg~I 5174, Na~I 5892) provide narrow
absorptions, while Diffuse Interstellar Bands (DIBs) are particularly 
prominent in the spectrum. To identify these, we use the catalog of 
\citet[][see also \url{http://leonid.arc.nasa.gov/DIBcatalog.html}]{jenet96}  
to synthesize a DIB absorption spectrum, combining Gaussian features at the cataloged
$\lambda_{\rm eff}$, kinematic and relative equivalent widths
(figure \ref{normspec}). Comparison with the companion
spectrum shows that many subtle broad features have a DIB origin. Note that the relative feature
strength shown here is only typical; substantial variation exists between various Galactic sight-lines.
The many DIB features make it particularly challenging to define the true photospheric continuum,
in turn this can subtly affect our line centroid estimates.

	The DIBs are believed to be generated by PAHs associated with interstellar dust. Thus
one expects a correlation between DIB strength and interstellar reddening from dust scattering.
\citet{koset13} lists fit linear relations for several strong DIBs of the form $EW = a \times E_{B-V} + b$.
These relations depend on whether the sight-line is `UV shielded' ($\zeta$ sight-line) or not
($\sigma$ sight-line), and the $\lambda$5797/$\lambda$5780 EW ratio is used to discriminate these
cases. Our value is $0.19\pm0.01$, below the 0.3 threshold, so we treat this as a $\sigma$ sight-line.  
In table \ref{DIB} 
we apply the tabulated coefficients to several well-measured DIBs to estimate the reddening. 
The inferred $E_{B-V}$ have a surprisingly large scatter, with a weighted average $E_{B-V} = 1.35 \pm 0.04$
(bootstrap error).
This value agrees very well with the estimate $E_{B-V} = 1.34 \pm 0.04$ determined by
\citet{napet11} from Str\"omgren photometry. It is similarly in reasonable accord with the
$N_H$ estimated from the X-ray spectrum. In turn, this agreement supports the photometric distance estimate
$d=5_{-2.1}^{+4.6}$\,kpc.

\begin{deluxetable}{rrrrr}[h!!]
\tablecaption{\label{DIB} Reddening from DIB Measurements}
\tablehead{
\colhead{DIB} & \colhead{a$^\dagger$} & \colhead{b$^\dagger$} &\colhead{EW} & $E_{\rm B-V}$\cr
\colhead{\AA } &\colhead{} & \colhead{ } &\colhead{ \AA} & \colhead{}
}
\startdata
5780&$0.506 \pm 0.050$ &$-0.014 \pm 0.032$  & $0.709 \pm 0.009$ & $1.43 \pm 0.16$ \cr
5797&$0.124 \pm 0.014$ &$-0.003 \pm 0.009$  & $0.134 \pm 0.006$ & $1.10 \pm 0.14$ \cr
6196&$0.048 \pm 0.004$ & $0.001 \pm 0.003$  & $0.063 \pm 0.005$ & $1.29 \pm 0.12$ \cr
6379&$0.057 \pm 0.012$ & $0.004 \pm 0.008$  & $0.119 \pm 0.005$ & $2.00 \pm 0.44$ \cr
6613&$0.192 \pm 0.026$ & $0.003 \pm 0.017$  & $0.225 \pm 0.006$ & $1.16 \pm 0.18$ \cr
6660&$0.014 \pm 0.004$ & $0.009 \pm 0.003$  & $0.043 \pm 0.004$ & $2.54 \pm 0.82$ \cr
\enddata
\tablenotetext{$^\dagger$}{Coefficients from \citet{koset13}.}
\end{deluxetable}

\subsection{Radial Velocity Measurements}

	To confirm the stability of our wavelength solutions, we measured the night sky 
lines $\lambda 5577.3387$\,\AA\ and $\lambda 6300.304$\,\AA\ in each processed spectrum, before
heliocentric corrections were applied.  The rms wavelength variations were $0.0106$\,\AA\ and $0.0124$\,\AA\, 
respectively, i.e. $\sigma_{RV} = 0.6 {\rm km\, s^{-1}}$.

	A second stability check was made measuring the very strong Na(D) interstellar doublet 
$\lambda 5889.951/5895.924$\,\AA\ in spectra corrected to the heliocenter. The line centroids display
mean redshifts of 2.88 and 2.85\,${\rm km\,s^{-1}}$, with spectrum-to-spectrum rms
variation of 0.88 and 0.74${\rm km\,s^{-1}}$, respectively. We conclude that our wavelength solutions are
good to better than 1 ${\rm km\,s^{-1}}$.

	To attempt to measure the photospheric velocity of the companion, we must follow the centroids
of the broad ($\sim 300 {\rm km\,s^{-1}}$) \ion{H}{1}, \ion{He}{1} and \ion{He}{2} lines, to a few ${\rm km\,s^{-1}}$. This
is challenging since the imperfect inter-order fluxing causes significant gradients and steps in the 
continuum, with a characteristic scale comparable to the typical 50\AA\ (3000 ${\rm km\,s^{-1}}$) order
width. Further, for several of the photospheric lines there are strong DIBs in the line wings, which
can perturb the continuum. To mitigate this latter effect, we divide the spectrum by the model DIB
spectrum. While imperfectly removing all DIBs, this does significantly
flatten the continuum in the region of the strongest features.

	All photospheric lines were fit with simple Gaussian profiles (test Voigt fitting showed negligible
Lorentzian wings). The fitting used the IRAF deblending routine, where the effective noise was determined for
each line by a measurement of the spectrum rms in the nearby, line-free continuum. The statistical errors on
the fit parameters (line centroid, strength and, optionally kinematic width) are provided by the routine,
based on Monte Carlo simulations. For the lower S/N lines the kinematic width was held fixed at the value
measured from the average spectrum (5-10\AA\ FWHM).

	We also attempted to determine the line centroids using the so-called ``mirroring'' method \citep{ps07}.
This allows one to separately probe the line wings and core. Although we can see that our lines are
significantly asymmetric, suggesting absorption in wind/outflow components, we lacked the S/N to measure
these structures accurately and compare them across lines and species. This prevented us from using the
line shapes to improve our estimate of the systematic (photospheric) radial velocity.

\begin{figure}[t!!]
\vskip 8.2truecm
\includegraphics{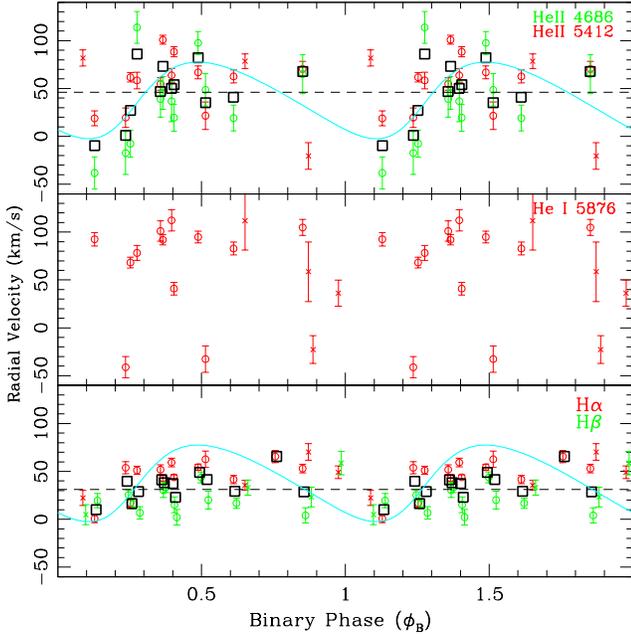}
\begin{center}
\caption{\label{RVlines} 
Radial Velocity measurements for three species. Points with circles are from high S/N spectra;
those with crosses from lower S/N data. Black squares show the mean velocity of the
two strongest lines for \ion{H}{1} and \ion{He}{2} for the high S/N spectra; large systematic scatter
is seen. The sample $K=40\,{\rm km\,s^{-1}}$ radial velocity curve plotted provides a 
reasonable match to the \ion{He}{2} average points; the H~I amplitude is $\sim 2.5\times$ smaller (see text).
}
\end{center}
\vskip -0.4truecm
\end{figure}

\section{Radial Velocity Variations}

	To test for velocity variation, we performed a weighted least-squares fit for the mean velocity 
of each species. Table 3 shows the velocities, errors, goodness of fit and the probability of constancy.
It is unclear how to compare mean velocities between lines, as the effective rest wavelengths depend on
the excitation. Also, the measured line centers are somewhat uncertain
due to continuum irregularities; for example \ion{He}{2} 6527 appears substantially redshifted,
due to a continuum gradient and superposed DIB. Finally, the lower excitation lines can be affected
by wind absorption and may be blue shifted with respect to \ion{He}{2} \citep{c05,set11}.

Overall, we infer a net systemic radial velocity 
of $\sim 40 {\rm km\,s^{-1}}$,
but a precise value requires cross-correlation with a simultaneously observed RV standard of similar
spectral class. The reduced $\chi^2$ values are, however, interesting. From the low
probabilities, we infer that the the strong lines do indeed show statistically significant variation.

\begin{deluxetable}{llcrr}[h!!]
\tablecaption{\label{RVtst} Test for Radial Velocity Variations}
\tablehead{
\colhead{Species} & \colhead{$\lambda_{\rm rest}$} & \colhead{$^\Gamma$} &\colhead{$\chi^2/$DoF} & \colhead{Log(Prob)}\cr
\colhead{} &\colhead{\AA\ in air} & \colhead{ ${\rm km\, s^{-1}}$} &\colhead{ } & \colhead{}
}
\startdata
H$\alpha$& 6562.808 & $43.0 \pm 1.1$ & $14.7$ & $-3.9$\cr
H$\beta$ & 4861.34  & $22.4 \pm 3.0$ & $ 3.6$ & $-1.2$\cr
H$\gamma$& 4340.48  & $25.5 \pm36.7$ & $ 3.3$ & $-1.2$\cr
\ion{He}{1}    & 5875.65  & $72.9 \pm 4.3$ & $20.8$ & $-5.3$\cr
\ion{He}{2}    & 4685.682 & $44.1 \pm20.9$ & $10.7$ & $-3.0$\cr
\ion{He}{2}    & 5411.52  & $63.3 \pm 3.3$ & $15.5$ & $-4.1$\cr
\ion{He}{2}    & 6527.099 &$125.8 \pm15.4$ & $ 5.6$ & $-1.7$\cr
\enddata
\end{deluxetable}

	However, there are certainly large systematic effects, and it is unclear how much of this variation
is associated with the photosphere radial velocity. To illustrate, we average the radial velocity
from the two strongest H~I lines (H$\alpha$ and H$\beta$) and two strongest \ion{He}{2} lines
($\lambda$4686 and $\lambda$5412) and plot the values for the highest quality (S/N$_{6000} > 20$) spectra
(Figure 3).
Even disegarding the mean radial velocity $\Gamma$ offset between lines, we see that the fluctuations are several
times the statistical error. Nevertheless, both sets of velocities show similar trends,
with relatively low values in the range $\phi_B=0.0-0.2$. Since $\phi_B=0$ is the $\gamma$-/X-ray 
light curve maximum, it is tempting to identify this excursion with periastron and interpret the 
radial velocity measurements as showing significant eccentricity. The two species \ion{H}{1} and \ion{He}{2}
show mean velocities $\Gamma_{\rm H~I} = 31.9{\rm km\,s^{-1}}$ and 
$\Gamma_{\rm He~II} =46.2{\rm km\,s^{-1}}$ and RMS dispersions $\sigma_{\rm H~I} = 11.1{\rm km\,s^{-1}}$
and  $\sigma_{\rm He~II} = 29.2{\rm km\,s^{-1}}$.

\begin{figure}[t!!]
\vskip 8.2truecm
\includegraphics{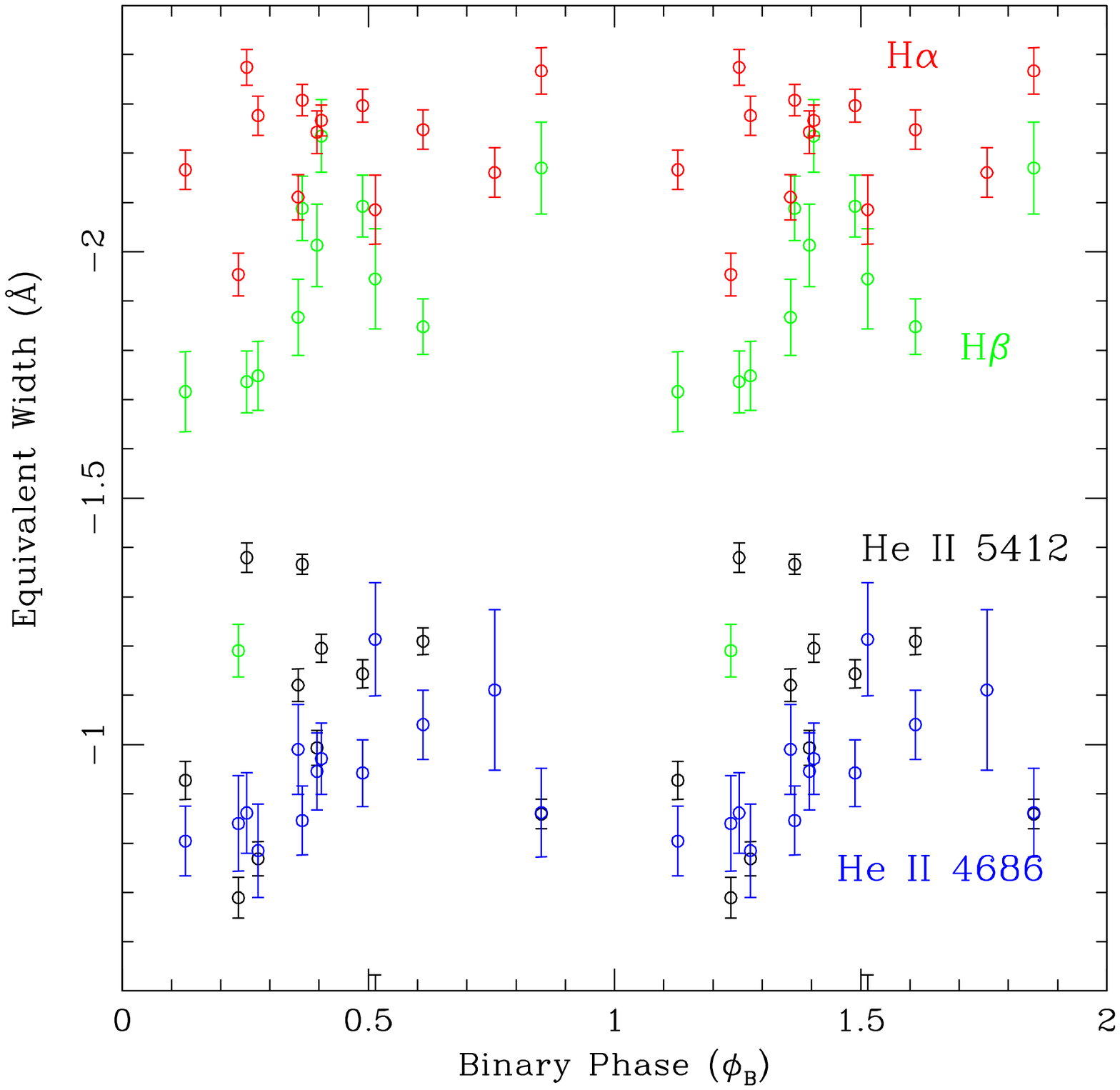}
\begin{center}
\caption{\label{EW} 
Equivalent width and its variation for strong absorption lines. H$\beta$ and \ion{He}{2} 4686
absorption appears to weaken near phase $\phi_B=0.2$.
}
\end{center}
\vskip -0.7truecm
\end{figure}

\subsection{Equivalent Width Variations}

	Following \citet{set11}, we checked the equivalent widths of our strongest absorption
lines. H$\alpha$, the strongest atomic absorption has an average equivalent width of
$-$2.2\AA. 
Like these authors we also see evidence for equivalent width variation through the orbit,
especially for H$\beta$ and \ion{He}{2} 4686 (Figure 4). However, in our case the minimum absorption strengths
appear to lie at $\phi_B \approx 0.2$, close to compact object superior conjunction for the
orbit suggested in figure \ref{RVlines} (i.e. when viewing the side of the companion away
from the compact object).  This might indicate denser wind flow through the L3 point.
Inspection of the strongest lines show variable wings, suggesting time or phase-dependent
views of such outflow.

	As for LS 5039, we find that our observed H$\alpha$ equivalent 
width, $-1.87$ to $-2.29$\AA\,, is smaller than the expected  $\approx -3.27$\AA\ photospheric width for this spectral 
class \citep{mcswainet04}. Scaling from those authors' estimates (i.e.
assuming similar companion mass and asymptotic wind velocity), we find that our lower observed residual 
absorption equivalent width implies a somewhat larger mass flux, namely ${\dot M} \approx 0.7-1.0 \times
10^{-7} M_\odot\,{\rm y^{-1}}$.

\subsection{Radial Velocity Model and Implied Mass}

	Labeling the primary (compact object) 1 and the visible secondary 2, we have a single line spectroscopic
binary, with radial velocity measurements constraining the mass function
$$
f(M_1) = \frac{M_1^3}{(M_1+M_2)^2} {\rm sin}^3 i = \frac{P}{2 \pi G} K_2^3 (1-e^2)^{\frac{3}{2}} 
$$
with $K_2 = \Delta v_2 / 2$ the radial velocity amplitude of the secondary and $i$ the unknown system inclination. 
Since ${\rm sin} i <1$ and $M_2>0$, $f(M_1)$ provides a lower bound to the mass of the primary.
\cite{c05} give the mass of an O6 V((f)) star as $20.0-26.4 M_\odot$.

	In view of the large and poorly understood systematics, one should not directly fit our radial
velocities. However, a basic model (figure \ref{RVlines}) gives an idea of the allowed amplitude.
Here a curve is plotted for $K=40 {\rm km\,s^{-1}}$, $\Gamma=40 {\rm km\,s^{-1}}$, $e=0.2$, and
longitude of ascending node $\omega=0.6\pi$. This provides a reasonable match to the range of the
\ion{He}{2} line; the \ion{H}{1} range is $K \approx 15 {\rm km\,s^{-1}}$. Larger amplitudes or eccentricities
provide a poor match to the data, so we take this as an upper limit on the allowed radial
velocity amplitude. 
\cite{set11} adopt the \ion{He}{2} radial velocity curve as minimizing wind contamination and
best following the photospheric motion. Their curve for \ion{H}{1} has a somewhat smaller amplitude than
that for \ion{He}{2}; for our data  the range of \ion{H}{1} velocities is $\sim 2.5\times$ smaller. Thus we follow
\cite{set11} in adopting the \ion{He}{2} velocities, but consider the possibility that these are systematics
dominated and that the stronger H~I lines reflect the true radial velocity amplitude.

\begin{figure}[b!!]
\vskip 8.3truecm
\includegraphics{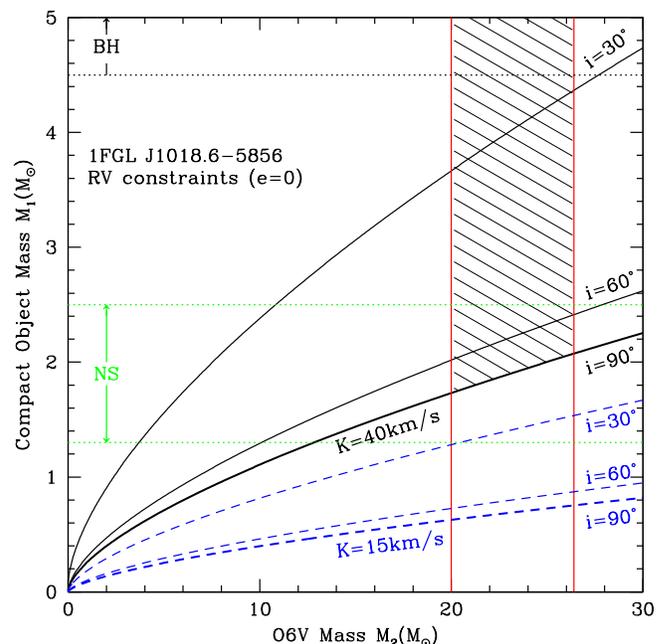}
\begin{center}
\caption{\label{m1m2} 
The binary mass-mass plane. For the \ion{He}{2}-estimated radial velocity amplitude, we see that a neutron star (NS)
solution requires small companion masses and/or large inclinations ($i>50^\circ$);
roughly half of the allowed parameter space is consistent with a black hole (BH), although typical BH masses require very
small inclinations. For the H~I amplitude, a neutron star is highly preferred, requiring
small inclinations to reach even $1.4M_\odot$.
}
\end{center}
\vskip -0.3truecm
\end{figure}

Figure \ref{m1m2} shows the corresponding allowed regions in the mass-mass plane for circular orbits. 
For the allowed range of eccentricity $M_1$ will be up to 7\% smaller. The solid lines show the 
$K=40 {\rm km\,s^{-1}}$ constraints suggested by \ion{He}{2}, the dashed lines show the range from
the Balmer lines.

\section{Discussion and Conclusions}

	Our attempt to measure the radial velocity curve of 1FGL 1018.6$-$5856 has been frustrated by a
large scatter, apparently due to systematic errors. In part these are certainly due to the imprecision
of the continuum calibration in our echelle spectra. In part they are also likely due to variable
absorption line components of a stellar wind unpredictably shifting the line centroids.

	However, we do see that the maximum radial velocity amplitude allowed by our data is
$\sim 40 {\rm km\,s^{-1}}$. If this is the true radial velocity amplitude, then we would require
a compact object mass $>1.7M_\odot$ and 50\% of the allowed parameter space has
$M_1 > 2.2 M_\odot$. Since the likely maximum neutron star mass is $\approx 2.5M_\odot$,
this would seem to suggest that the $\gamma$-ray source is a black hole. However, one should 
compare with the masses of other stellar mass black holes. \citet{faet11} find that
at 90\% confidence the minimum $M_{BH}$ in stellar mass binaries is $>4.5M_\odot$. Although
they do not separately evaluate this for the high mass systems, like 1FGL 1018.6$-$5856,
their figure 2 indicates most probable masses $10-20M_\odot$ and 90\% lower mass limits $>10M_\odot$.
From figure \ref{m1m2} we see that we require $i<30^\circ$ for $M_1>4.5M_\odot$ ({\it a priori} probability 13\%),
while for $M_1>10M_\odot$ we require $i<14^\circ$ (probability 3\%). Thus the likelihood of 
having a typical stellar black hole mass seems modest. When we consider that systematic
errors may be contributing to the \ion{He}{2} radial velocity width and that the 
H~I radial velocity amplitude may be closer to the intrinsic amplitude, this
only allows $M_1$ to exceed $4.5M_\odot$ for $i<11^\circ$
(probability 2\%). Thus we conclude that the modest radial velocity amplitude observed for
1FGL 1018.6$-$5856 prefers a neutron star. However, these statistical statements do not
exclude either option and a clean measurement of the companion radial velocity would
help the situation greatly.

	The way forward on determining the mass of 1FGL J1018.6$-$5856 doubtless starts with improved
spectroscopy. Our data show that the photospheric lines are quite broad, and likely variable. Accordingly
a future campaign on small- medium class telescopes should obtain long-slit, moderate resolution
($R\sim 3000-6000$) spectra covering 4000-5500\AA, with a collection
of several template O6 V radial velocity comparisons. Queue/service observing is almost certainly
required in view of the large $P_b$. Higher resolution echelle data from a large telescope can be of use,
but will require higher S/N to allow deconvolution of the line shape and isolation of the
systematic radial velocity. If the minimum mass exceeds $3M_\odot$, we can be confident of a black
hole primary. But in the more likely event that lower masses are allowed, the inclination becomes
critical and one will require very
high precision photometry to constrain the small ellipsoidal variations, or constraints from
light curve fitting at other wavebands.

\acknowledgements

We thank Robin Corbet, Malcolm Coe, Richard Dubois and Matthew Kerr for assistance with original
proposal. HongJun An helped with useful discussions about the orbit ephemeris and Matt Giguere 
assisted with the CHIRON data acquisition.

	This project made use of data obtained at the CTIO 1.5\,m telescope run by the SMARTS
consortium. CTIO is operated by the Association of 
Universities for Research in Astronomy, under contract with the National Science Foundation.
This work was supported in part by NASA grants NNX11AO44G and NAS5-00147.

\end{document}